*Original and Official Version*
*15/12/2023*

# Literature Review of Mixed Reality Research

Aizierjiang Aiersilan[1,*]

[1]Faculty of Science and Technology, University of Macau, Macau, China

**Abstract.** In the global context, while mixed reality has been an emerging concept for years, recent technological and scientific advancements have now made it poised to revolutionize industries and daily life by offering enhanced functionalities and improved services. Besides reviewing the highly cited papers in the last 20 years among over a thousand research papers on mixed reality, this systematic review provides the state-of-the-art applications and utilities of the mixed reality by primarily scrutinizing the associated papers in 2022 and 2023. Focusing on the potentials that this technology have in providing digitally supported simulations and other utilities in the era of large language models, highlighting the potential and limitations of the innovative solutions and also bringing focus to emerging research directions, such as telemedicine, remote control and optimization of direct volume rendering. The paper's associated repository is publicly accessible at [https://aizierjiang.github.io/mr](https://aizierjiang.github.io/mr).

## 1 Introduction

The last few years have seen the rapid development of Metaverse-related[1] technology, particularly with the emergence of Large Language Models (LLMs) like ChatGPT[2], LLMA[3] and PaLM[4]. These large language models hold the potential to significantly enhance human-computer interaction, propelling it to new levels of sophistication and capability. As an important stream of future research in human-computer interaction (HCI), mixed reality (MR) demonstrates its full potential in leveraging LLMs within the modern era. Moreover, concurrent with hardware enhancements, various innovative and cutting-edge MR devices with remarkable advancements such as Microsoft HoloLens series, Magic Leap series, Meta Quest series, Lenovo ThinkReality VRX, Varjo XR, Google Glass and Apple Vision Pro, etc., have emerged. These are all evident facts depicting an extensive potential in MR application and research. However, based on the review towards recent papers published in the last 2 decades including the year 2023, I found that the research regarding MR is declining since 2020. The annual publication rate (Figure 1) for MR research peaked in 2020 on Google Scholar, coinciding with the introduction of the Metaverse concept during the COVID-19 pandemic, followed by a subsequent decline in 2023, nearly aligning with the publication rate observed in 2016. Does this imply that MR research is experiencing a slowdown, reaching a technological bottleneck? If so, what potential applications could expedite academic research? Based on such questions, this paper discusses highly cited MR research papers and recently published scholarly works focusing on MR.

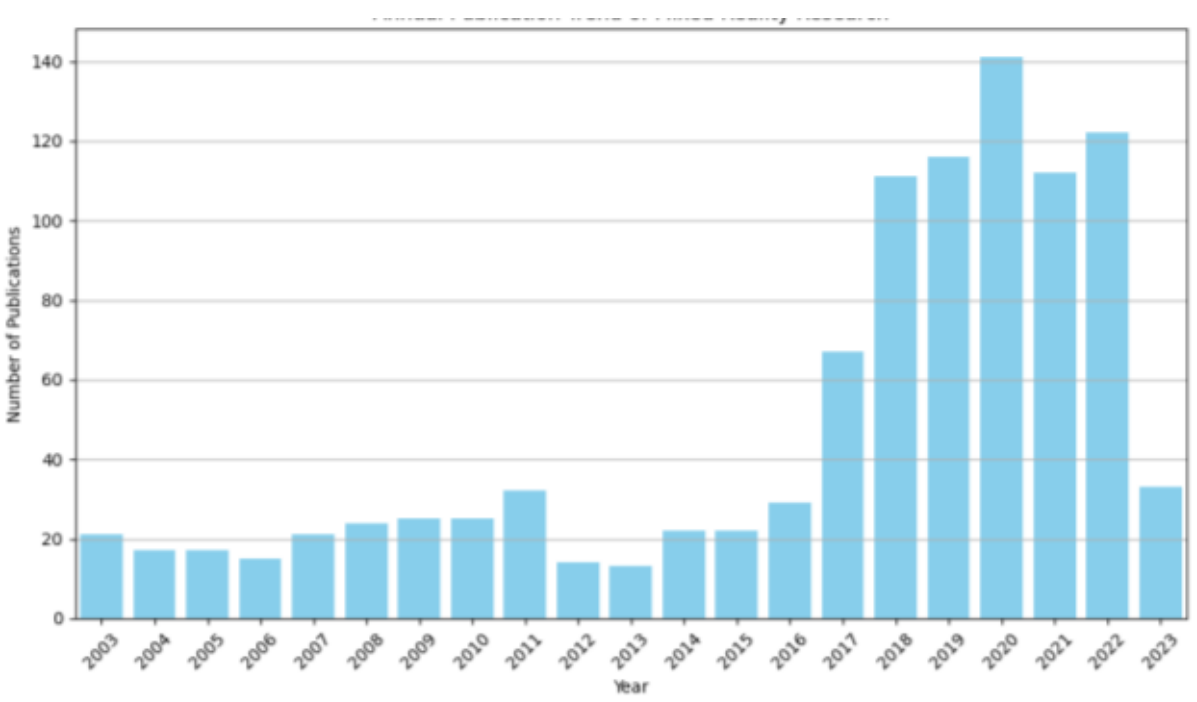

Figure 1: Annual publication trend of MR-related research from 2003 to 2023

[5] introduces the definitions of MR and explores the scope of MR applications. [6] enriches the study on the utility of MR from different aspects in our life. [7] lists an array of MR devices. Most of the previous surveys or reviews predominantly showcased advancements in both hardware and software within the realm of MR, highlighting the prevailing challenges encountered in the industry. However, there remains a dearth of studies that articulate how MR, in turn, expedites the advancement of research pertaining to its underlying technologies and theories. Rather than reiterating the recognized definitions of MR and listing features of prevalent MR instruments, this paper directly focus on the highly cited and recently published academic papers on MR, aiming to uncover potential benefits of MR for future research.

## 2 Search Strategy

Utilizing the software *Publish or Perish* [8], an efficient search tool for researchers, my initial step involved keyword-based searches spanning from 2003 to 2023. I initiated my study from the year 2003 to track the longitudinal development trends within the past two decades. It was intriguing to observe that despite the absence of much of the current hardware ten years prior to my starting point, research in this domain has continued to persist. Utilizing specific title keywords in Google Scholar, I retrieved pertinent papers, enabling a comprehensive overview of the field's evolution over the course of twenty years. Sorting these papers by their citation count facilitated my comparative analysis. Through sorting by their citation count, I was surprised to find that the highly cited papers predominantly originated from diverse fields such as education, medicine, business, and management. Upon closer examination of each collected paper, they focus more on conceptual designs and definitions rather than delving into the technical aspects of MR. To refine my study's scope and prevent redundancy, I manually curated highly cited papers within computer graphics, computer vision, communication and networks, and other areas like education and medicine, excluding duplicate work that had already been covered by other surveys and literature reviews.

Among the selected 100 highly cited papers from 2017, a period marked by the escalating research interest in MR, the majority focus on discussing the application of MR in various fields rather than delving into the in-depth research underlying its advanced technologies like spatial computing, graphics and rendering, computer vision, pattern recognition, and other pertinent components, indicating that MR is actually just a concept behind the customer-oriented applications rather than a specific technology or a set of theories that can be deeply researched by the scientists and scholars (Figure 2). Identifying the primary focus of modern academic research in computer science from [9] is straightforward. Research in computer science often focus on a specific topic at least in the stage of *Technologies* and high level of research cannot be conducted without certain abstraction into the stage of *Theories*. It is most common that the industry proposes concepts and designs of the customer-oriented applications from the products, which later be treated as a demand to the academics, and the researchers dig deeper into the technologies behind, and scientists dig much deeper to explore their theories and make optimizations.

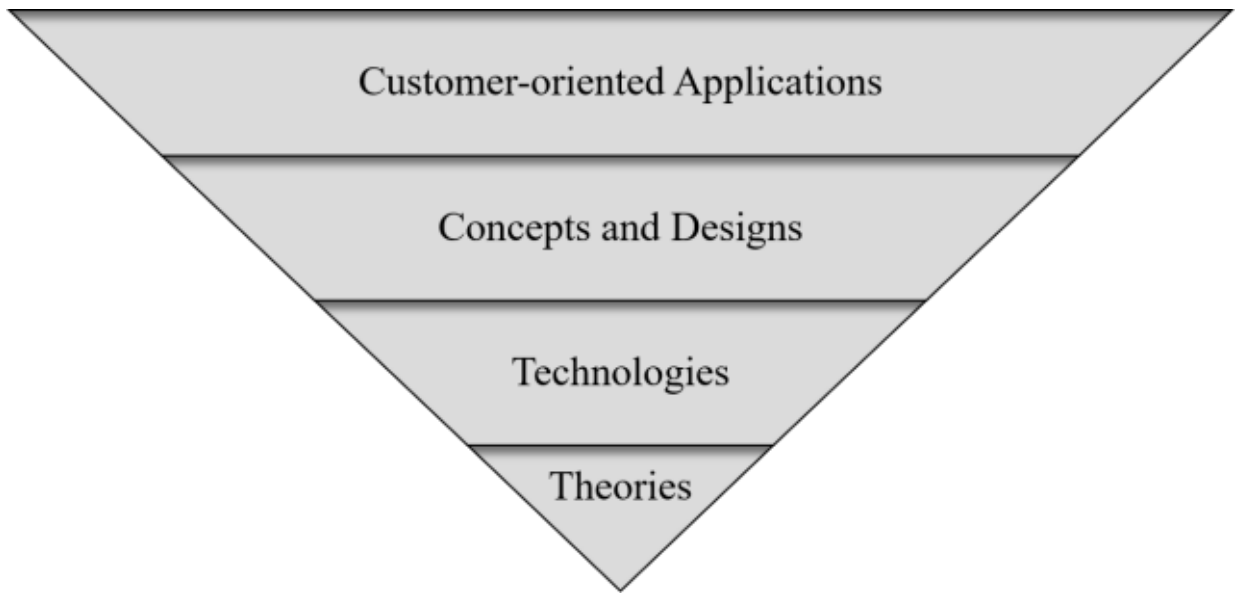

Figure 2 : The hierarchical model of a specific high-tech product. From *Customer-oriented Applications*, which are directly accessible for end-users, to the core principles and *Theories* primarily investigated by scientists in relevant fields, the entire spectrum plays a crucial role in shaping a comprehensive high-tech product.

## 3 Literature Review

Since 2017, the predominant choice in academia of MR has been the utilization of HoloLens as the primary device to materialize research concepts and ideas, establishing it as one of the renowned MR devices widely recognized within both academic and industrial spheres. Most of the researches conducting experiments and developing applications in recent years are from game industry and medical science, surpassing involvement in other domains. It is clear that substantial opportunities exist for immersive simulations within gaming content as well as clinical and medical cases.

Remarkable advantages have already been identified from using the HoloLens for medical use, from training in anatomy and diagnostics to acute and critical patient care, such as for visualizing organs prior to surgery, teaching dental students, and in pathology education, etc. In [10], researchers utilized images gained from standard neuro-navigation magnetic resonance imaging for intra-operative planning, which were subsequently uploaded onto the HoloLens 2 system to generate a 3D image. This enabled surgeons to overlay a 3D rendered image onto the patient's anatomy, thereby mitigating the limitations of conventional neuro-navigation systems. In [11], the researchers employed HoloLens 2 to facilitate the visualization of computerized tomography scan data and Digital Imaging and Communications in Medicine files. As rendering extensive volumes poses challenges on mobile devices due to limited memory capacity and device constraints, an Empty Space Skipping optimization algorithm to enable high-quality rendering of large models was integrated on the HoloLens. The crux of the method is to generate a ray for each voxel, which originates from the center of the camera and passes through the imaginary image plane between the camera and the volume used for rendering.

An imposter technique utilizing 2D billboards to depict 3D objects is employed and the Empty Space Skipping algorithm that enhances the efficiency of ray marching through vacant or transparent regions is implemented thereafter. Eventually, a pre-calculated distance map to retain the shortest distance from an opaque voxel to an empty voxel is generated. The outcome of this approach reportedly achieved approximately tenfold frames per second enhancement on a high-quality model compared to the non-optimized version.

Besides medical imaging, there are also great amount of applications of MR devices in mental science and anatomy. The researchers of [12] presents a cutting-edge educational tool to control phobia-related dread. It is widely recognized that instructing an individual suffering from a phobia to identify and control their fear is a pivotal aspect of the therapy regimen. By employing MR software, the patient can be exposed to a range of distressing scenarios associated with their phobia and observe the ensuing effects on their emotional state. This facilitates a more accurate evaluation of the phobia's severity and subsequently enhances coaching strategies for the patient to manage their fear. The researchers of the paper used HoloLens 2 to display and render the 3D models into participant's real environment, meaning that the participant can be in contact with their phobia directly in an environment they are familiar with (e.g. their apartment, room, etc). For individuals experiencing arachnophobia, spiders could be projected onto the floor and walls of their room through the glasses worn on their head. In [13], a systematic review is provided with the state-of-the-art applications of the MR in a medical and healthcare context. Focusing on the potential of this technology in digitally supported clinical care, studies showcasing the applicability and feasibility of MR in medical and healthcare settings are mainly considered. It highlights the potential and limitations of the current MR-based innovative solutions and bring focus to emerging research topics, such as telemedicine, remote control and motor rehabilitation.

While aforementioned studies primarily explore the utilization of MR technology for simulation purposes, [14] provides a short overview of the current state focusing on medical applications using smart glasses and evaluated the HoloLens glasses regarding latency and data rates by using WiFi and the 5G campus network. To quantify both the uplink and downlink speeds, alongside response time, a "speedtest" is conducted within a browser window. The Remote Access application enabled share communication between the HoloLens and tablet, permitting calls to be initiated solely from the device, thereby streaming its picture. Both devices are connected to the same base station, using a non-standalone core network for communication. The study highlights the potential use of the MR in educational, training, and teleassistance applications but pinpoint challenges regarding latency, network stability, ergonomics, and privacy. As 5G ensures the better network condition for the low-friction plus immersive communications among 3D contents, [15] proposes a prototype implementation of real-time point cloud streaming system (Figure 3) and showed live-action demonstrations of live point cloud streaming using MR devices.

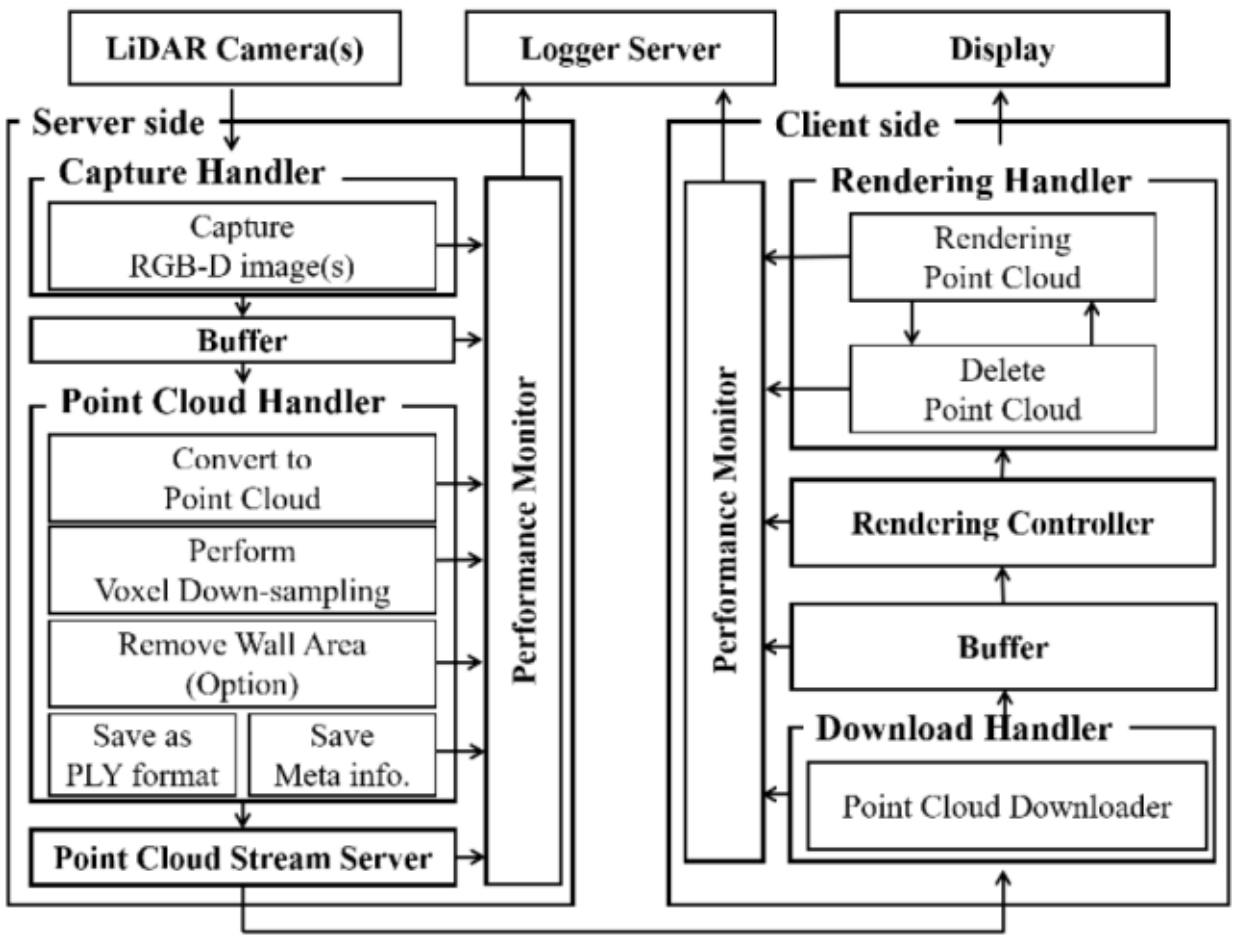

Figure 3: System architecture of point cloud streaming system

Their presentation on network delay illustrated the prototype system's potential capacity to support uninterrupted live streaming without any interruptions in point cloud playback.

Remote collaboration's increasing significance in demonstrating practicality necessitates high-quality rendering, especially in remote environments. In [16], the problem of latency is also studied under the case of remote collaboration. It introduces a remote collaboration technique using holographic avatars for real-time movement transfer, facilitating immersive interactions with 3D visualized data among multiple remote users (Figure 4).

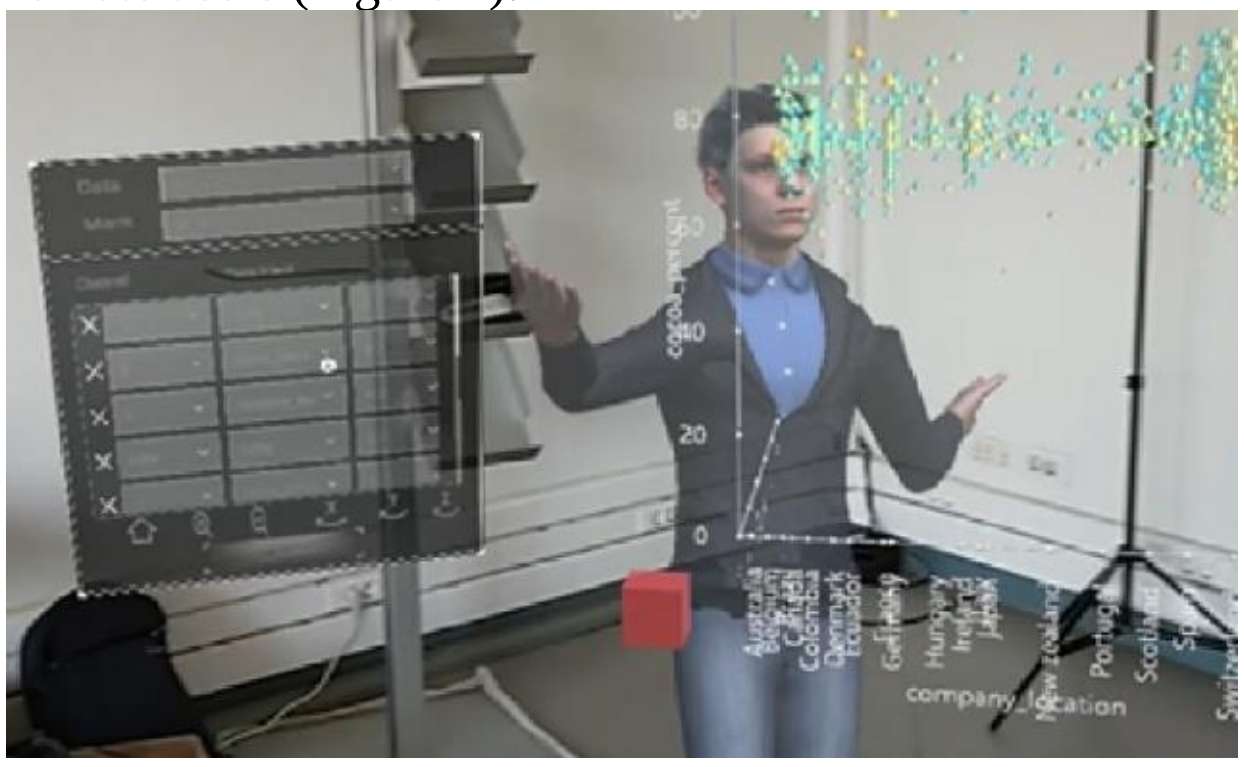

Figure 4: The efficient remote collaboration via MR

[17] proposes an optimized remote rendering enabled extended reality (XR) system (Figure 5) that presents the 3D city model of New York City on HoloLens.

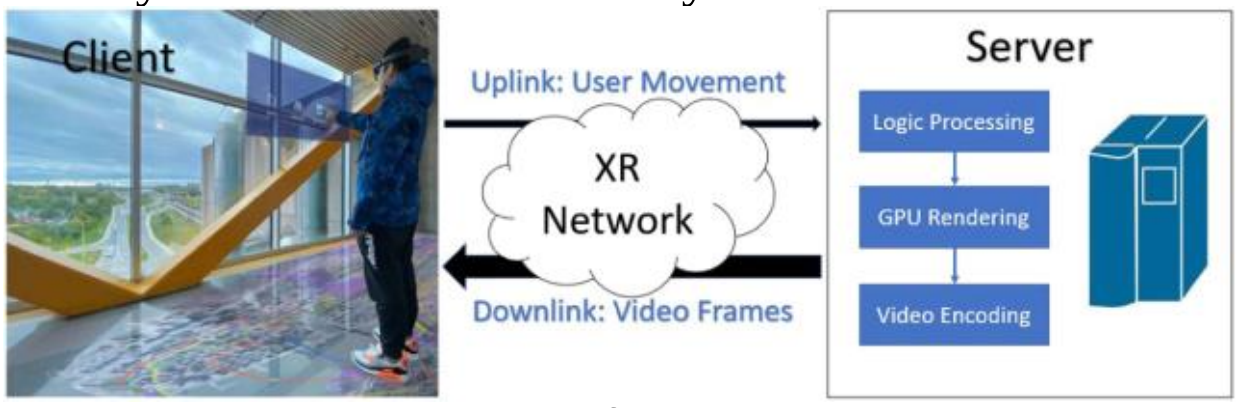

Figure 5: The architecture of the system

The findings suggest that remote rendering exhibits superior performance compared to local rendering for the model, resulting in a substantial enhancement in the average Quality of Experience (QoE) by a minimum of 21%. Their QoE is quantified utilizing the formula provided below:

$$\mathrm{QoE} = \sum_{n=1}^{N} q(F_n) + \sum_{n=1}^{N} p(R_n) - u \sum_{n=1}^{N} g(L_n)$$

In the given equations, $q(F_n)$ denotes user satisfaction with video smoothness at time window n based on average frame rate ($F_n$), while $p(R_n)$ signifies satisfaction with video clarity at time window n based on average resolution ($R_n$). Also, $g(L_n)$ penalizes latency between transmitting XR device pose data to the remote server and displaying the corresponding video frame, utilizing average latency ($L_n$) at time window n.

In [18], researchers propose a solution based on the use of a magnifier placed in front of a commercial optical see-through head-mounted display to mitigate the focus rivalry conflict between the virtual content and the real environment in the peripersonal space in order to leverage the profitable use of the current generation of optical see-through headsets for high-precision manual tasks. The preliminary user test findings indicate potential success in consistently mitigating focus rivalry, but additional research is required to address motion-parallax compensation and magnification mismatch between virtual content and real targets projected at the display focal distance.
Aside from research in medical science or surgery, many studies and applications exist within the realms of game development and real-time rendering. In [19], more rendering optimization details about MR is studied in algorithmic perspective. As direct volume rendering is a standard technique for visualizing scientific volumetric data in three-dimension, the researchers of the paper explored a key requirement of rendering latency capability for MR head-mounted displays by proposing a benchmark application with 5 volumes and 30 rendering parameter variations (Figure 6).

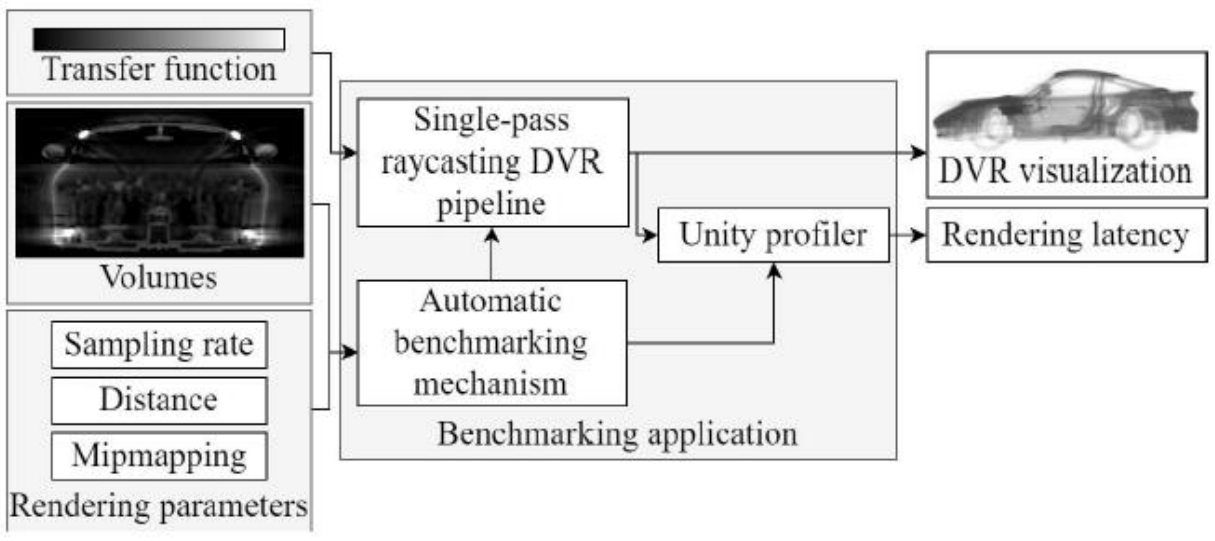

Figure 6: The evaluation process for ray-casting algorithm

The study tests three rendering parameters: sampling rate, rendering distance, and mipmapping, creating 30 variations (5 sampling rates × 3 rendering distances × 2 mipmapping). Using HoloLens, rendering latency is evaluated across 100 frames during random rotations of the volume. Lower-resolution volumes shows better average rendering latency. Mipmapping notably improves average rendering latency on HoloLens, categorizing more direct volume renderings as optimal, practical, or minimal. Higher sampling rates correlates positively with rendering latency, while rendering distance had a negative correlation. Overall, the study suggests that enabling mipmapping on HoloLens enables interactive direct volume rendering with standard resolution volumes.
Ultimately, the primary constraint observed in MR devices, as highlighted in [20], is their limitation (Figure 7) in outdoor settings due to their restricted operational distances, which typically extend up to 3.5 meters.

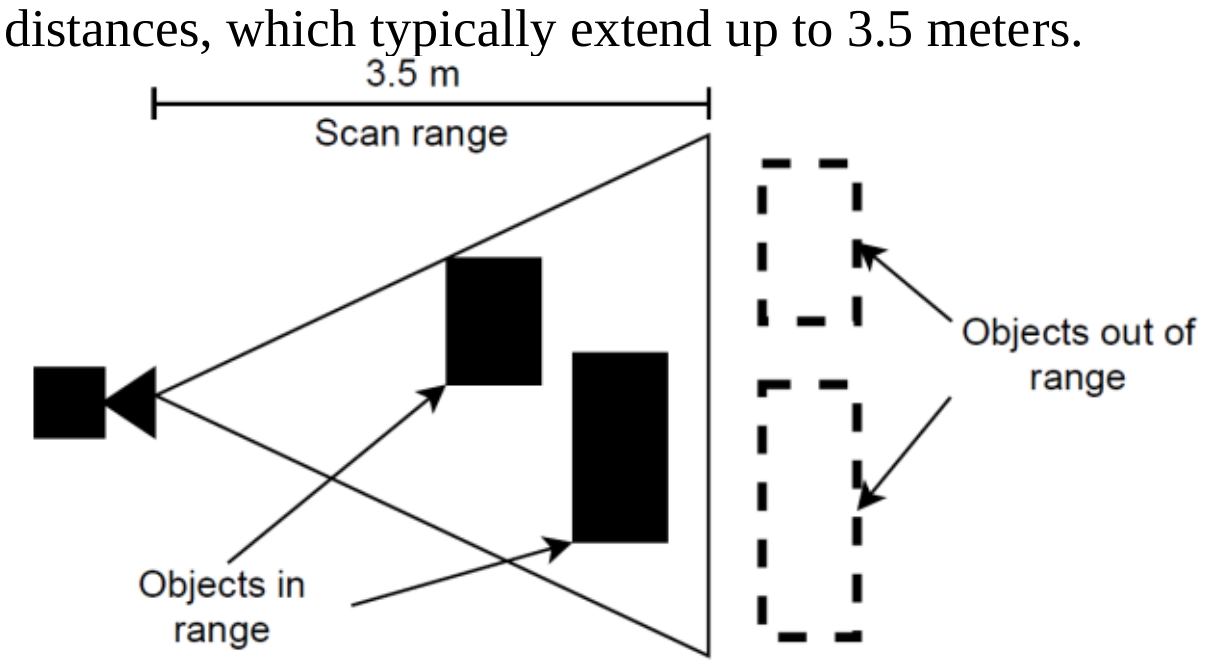

Figure 7: Scanning limitations of current MR devices.
Within the context of MR, current head-mounted displays encounter limitations by being unable to link their local coordinate systems with a global position and orientation, hindering their multi-user functionality, particularly in outdoor settings where spatial information sharing is challenging. To address this, the researchers introduced an approach using the CityJSON format—a JavaScript Object Notation-based framework for 3D models within customizable coordinate systems—to share geo-referenced 3D geometries across multiple devices, primarily for occlusion purposes in geospatial MR visualization (Figure 8).

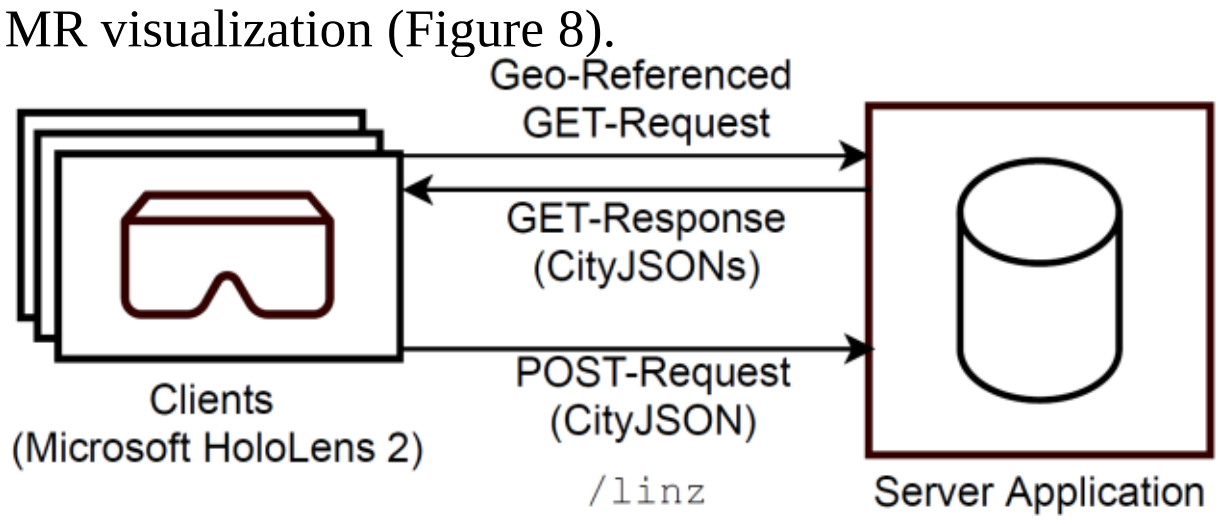

Figure 8: System architecture

Besides the traditional aspects of MR, some attempts on integrating generative AI to MR have appeared in recent studies. In [21], a simple framework of AI-based simulation for autonomous driving in vehicular MR under the horizon of Metaverses is proposed. Though the proposed framework seems just follow the trend of fancy concepts like Metaverse and has little contribution to MR itself, its framework states the communication challenges in MR. [22] surveyed on the remote assistance and training applications in MR, while [23] introduces using replicas to improve remote collaboration in MR. The highly cited paper in 2023 about MR is also a survey clarifying the differences between virtual reality, augmented reality and MR.

## 4 Gaps and Controversies

The exploration of MR research underscores a notable disparity in approach among researchers. While those in the computer science domain primarily concentrate on the technological aspects underpinning MR applications, counterparts in other fields actively employ MR as an emerging paradigm to elevate their areas of inquiry. A discernible gap exists between researchers specializing in computer graphics, computer vision, networks, and the domain of MR itself. The majority of researchers in these fields have not directed their investigations towards MR applications or sought solutions within the realm of MR, consequently impeding the advancement of MR technology to a certain extent. Additionally, it is unrealistic to expect researchers outside the computer science domain to address the complexities inherent in MR-related technologies.
Regrettably, despite my exhaustive investigation, a noticeable gap persists in the integration of LLMs to MR by the end of 2023. While LLMs have undergone rapid development and are widely adopted across various research domains, the full amalgamation of LLMs with MR research remains elusive. The diminutive size of these devices poses challenges for running LLMs; however, opportunities exist to optimize the network efficiency of MR devices for the seamless integration of cutting-edge applications like AI consultancy, AI assistant and AI 3D virtual friend. Should LLM-based applications be successfully integrated into MR systems, it becomes evident that the Metaverse is within imminent reach.

## 5 Conclusion

My analysis involves an intricate examination of both recent works over the last two years and the overarching evolution of MR over a span of two decades. Globally, MR has displayed a gradual emergence, yet recent strides in technology and science have propelled it towards transformative potential across various industries and everyday life. This comprehensive review thoroughly investigates the landscape of MR by methodically assessing highly cited papers from the past two decades, encompassing a broad spectrum of research publications on this subject. With a primary focus on scrutinizing papers released in 2022 and 2023, this study explores the contemporary applications and functionalities of MR. As of present, the integration of LLMs into MR remains absent. However, I anticipate a promising future trajectory by leveraging LLMs to enhance MR capabilities and augment its effectiveness. The search strategy and the metrics I used when making this literature review is open source and publicly accessible through the project page https://aizierjiang.github.io/mr.

## References

1. Mystakidis, Stylianos. "Metaverse." Encyclopedia 2.1 (2022): 486-497.
2. Ray, Partha Pratim. "ChatGPT: A comprehensive review on background, applications, key challenges, bias, ethics, limitations and future scope." Internet of Things and Cyber-Physical Systems (2023).
3. Touvron, Hugo, et al. "Llama: Open and efficient foundation language models." arXiv preprint arXiv:2302.13971 (2023).
4. Chowdhery, Aakanksha, et al. "Palm: Scaling language modeling with pathways." Journal of Machine Learning Research 24.240 (2023): 1-113.
5. Speicher, Maximilian, Brian D. Hall, and Michael Nebeling. "What is mixed reality?." Proceedings of the 2019 CHI conference on human factors in computing systems. 2019.
6. Hughes, Charles E., et al. "Mixed reality in education, entertainment, and training." IEEE computer graphics and applications 25.6 (2005): 24-30.
7. Verhey, Jens T., et al. "Virtual, augmented, and mixed reality applications in orthopedic surgery." The International Journal of Medical Robotics and Computer Assisted Surgery 16.2 (2020): e2067.
8. Harzing, Anne-Wil. The publish or perish book. Melbourne, Australia: Tarma Software Research Pty Limited, 2010.
9. Dodig-Crnkovic, Gordana. "Scientific methods in computer science." Proceedings of the Conference for the Promotion of Research in IT at New Universities and at University Colleges in Sweden, Skövde, Suecia. sn, 2002.
10. Jain, Swati, et al. "Use of mixed reality in neurosurgery OR: first experience with HoloLens 2." Brain Tumor Research and Treatment 10.Suppl (2022).
11. Cetinsaya, Berk, Carsten Neumann, and Dirk Reiners. "Using Direct Volume Rendering for Augmented Reality in Resource-constrained Platforms." 2022 IEEE Conference on Virtual Reality and 3D User Interfaces Abstracts and Workshops (VRW). IEEE, 2022.
12. Janecký, Dominik, Erik Kučera, and Oto Haffner. "HoloLens 2 and Virtual Reality as Methods for Managing Phobias." 2022 Cybernetics & Informatics (K&I). IEEE, 2022.
13. Palumbo, Arrigo. "Microsoft Hololens 2 in medical and healthcare context: state of the art and future prospects." Sensors 22.20 (2022): 7709.
14. Wersényi, György. "Evaluation of the HoloLens for Medical Applications Using 5G-connected Mobile Devices." INFOCOMMUNICATIONS JOURNAL 14.4 (2022): 11-17.
15. Chujo, Yumeka, Kenji Kanai, and Jiro Katto. "Implementation and Demonstration of Real-time Point Cloud Streaming System using HoloLens." 2022 IEEE International Conference on Consumer Electronics-Taiwan. IEEE, 2022.
16. Farouk, Passant, Nourhan Faransawy, and Nada Sharaf. "Using HoloLens for Remote Collaboration in Extended Data Visualization." 2022 26th International Conference Information Visualisation (IV). IEEE, 2022.
17. Long, Zijian, Haiwei Dong, and Abdulmotaleb El Saddik. "Interacting with New York city data by HoloLens through Remote Rendering." IEEE Consumer Electronics Magazine 11.5 (2022): 64-72.
18. Condino, Sara, et al. "How to Mitigate Perceptual Limits of OST Display for Guiding Manual Tasks: a Proof of Concept Study with Microsoft HoloLens." 2022 IEEE International Conference on Metrology for Extended Reality, Artificial Intelligence and Neural Engineering (MetroXRAINE). IEEE, 2022.
19. Jung, Hoijoon, Younhyun Jung, and Jinman Kim. "Understanding the Capabilities of the HoloLens 1 and 2 in a Mixed Reality Environment for Direct Volume Rendering with a Ray-casting Algorithm." 2022 IEEE Conference on Virtual Reality and 3D User Interfaces Abstracts and Workshops (VRW). IEEE, 2022.
20. Praschl, Christoph, and Oliver Krauss. "Geo-Referenced Occlusion Models for Mixed Reality Applications using the Microsoft HoloLens." VISIGRAPP (3: IVAPP). 2022.

21. Xu, Minrui, et al. "Generative AI-empowered simulation for autonomous driving in vehicular mixed reality metaverses." arXiv preprint arXiv:2302.08418 (2023).
22. Fidalgo, Catarina G., et al. "A Survey on Remote Assistance and Training in Mixed Reality Environments." IEEE Transactions on Visualization and Computer Graphics 29.5 (2023): 2291-2303.
23. Tian, Huayuan, et al. "Using Virtual Replicas to Improve Mixed Reality Remote Collaboration." IEEE Transactions on Visualization and Computer Graphics 29.5 (2023): 2785-2795.